\documentclass[sigconf, nonacm]{acmart}
\usepackage[a-2b]{pdfx}

\newcommand\vldbdoi{10.14778/3827998.3828032}
\newcommand\vldbpages{4276 - 4289}
\newcommand\vldbvolume{19}
\newcommand\vldbissue{12}
\newcommand\vldbyear{2026}
\newcommand\vldbauthors{\authors}
\newcommand\vldbtitle{\shorttitle} 
\newcommand\vldbavailabilityurl{}
\newcommand\vldbpagestyle{empty} 

\usepackage{seqsplit}

\usepackage[ruled,linesnumbered,vlined]{algorithm2e}

\usepackage{enumitem}
\setlist[itemize]{leftmargin=1.5em}
\setlist[enumerate]{leftmargin=1.5em}

\usepackage{listings}
\usepackage{xcolor}

\setlist[itemize]{leftmargin=1.5em, topsep=2pt, parsep=1pt, itemsep=2pt}

\definecolor{sqlkeyword}{RGB}{0,0,180}
\definecolor{sqlcomment}{RGB}{100,100,100}
\definecolor{sqlstring}{RGB}{170,55,55}
\definecolor{sqltype}{RGB}{0,128,0}
\definecolor{sqlbg}{RGB}{248,248,248}

\lstdefinestyle{sqlstyle}{
  language=SQL,
  basicstyle=\ttfamily\scriptsize,
  keywordstyle=\color{sqlkeyword}\bfseries,
  commentstyle=\color{sqlcomment},
  stringstyle=\color{sqlstring},
  showstringspaces=false,
  breaklines=false,
  frame=single,
  framesep=4pt,
  xleftmargin=4pt,
  xrightmargin=4pt,
  aboveskip=8pt,
  belowskip=8pt,
  columns=flexible,
  backgroundcolor=\color{sqlbg},
  rulecolor=\color{black!30},
  morekeywords={STRING, STRUCT, NOT, NULL, TABLE, CREATE},
  morekeywords=[2]{STRING, STRUCT, BIGINT, DOUBLE, INT},
  keywordstyle=[2]{\color{sqltype}\bfseries},
  literate=
    {<Type1>}{{{\color{sqltype}<Type1>}}}{7}
    {<TypeN>}{{{\color{sqltype}<TypeN>}}}{7}
    {<...>}{{{\color{sqltype}<...>}}}{5}
}

\usepackage{microtype}

\begin{document}

\title{OmniTable: A Unified Wide-Table System for Petabyte-Scale LLM Data Curation and Exploration}

\author{Yuzhuo Fu}
\email{fuyuzhuo.fyz@antgroup.com}
\affiliation{\institution{AntGroup}}

\author{Xiangchun Wang}
\email{wangxiangchun.wxc@antgroup.com}
\affiliation{\institution{AntGroup}}

\author{Chao Huang}
\email{zhonggong.hc@antgroup.com}
\affiliation{\institution{AntGroup}}

\author{Liyi Wang}
\email{banzhu.wly@antgroup.com}
\affiliation{\institution{AntGroup}}

\author{Binwei Zeng}
\email{binwei.zbw@antgroup.com}
\affiliation{\institution{AntGroup}}

\author{Yuhan Wang}
\email{heqi.wyh@antgroup.com}
\affiliation{\institution{AntGroup}}

\author{Taotao Nie}
\email{nietaotao.ntt@antgroup.com}
\affiliation{\institution{AntGroup}}

\author{Dongke Hu}
\email{dongke.hudk@antgroup.com}
\affiliation{\institution{AntGroup}}

\author{Wang Hong}
\email{hongwang.hw@antgroup.com}
\affiliation{\institution{AntGroup}}

\author{Jiayi Wang}
\email{jiaer.wjy@antgroup.com}
\affiliation{\institution{AntGroup}}

\author{Wenwen Cui}
\email{cww516675@antgroup.com}
\affiliation{\institution{AntGroup}}

\author{Zhuyan Zhou}
\email{zhouzhuyan.zzy@antgroup.com}
\affiliation{\institution{AntGroup}}

\author{Yushun Guo}
\email{guoyushun.gys@antgroup.com}
\affiliation{\institution{AntGroup}}

\author{Yuhan Xing}
\email{xingyuhan.xyh@antgroup.com}
\affiliation{\institution{AntGroup}}

\author{Jiaxin Lian}
\email{lianjiaxin.ljx@antgroup.com}
\affiliation{\institution{AntGroup}}

\author{Peng Lin}
\email{peng.linlp@antgroup.com}
\affiliation{\institution{AntGroup}}

\author{Qing Cui}
\email{cuiqing.cq@antgroup.com}
\affiliation{\institution{AntGroup}}

\author{Wenhui Shi}
\email{yushun.swh@antgroup.com}
\affiliation{\institution{AntGroup}}

\author{Jun Zhou}
\authornote{Corresponding author.}
\email{jun.zhoujun@antgroup.com}
\affiliation{\institution{AntGroup}}

\begin{abstract}
Data curation is a critical bottleneck in industrial-grade LLM development, where petabyte-scale unstructured corpora are scattered across hundreds of physical tables, feature engineering relies on manual, table-centric pipeline orchestration, and data lineage is largely absent. We present OmniTable as an architecture blueprint for a unified wide-table layer built on Logical Unification, Physical Separation, targeting petabyte-scale LLM data curation and exploration. OmniTable makes four contributions: (1) a unified wide-table abstraction that consolidates multi-source heterogeneous data and thousands of derived features under a single logical schema via logical-physical mapping; (2) declarative feature lifecycle management that automates dependency resolution, execution planning, operator fusion, and lineage tracking, replacing manual pipeline orchestration with a "declare-and-execute" paradigm; (3) an adaptive execution engine with autonomous governance that achieves stable PB-scale feature backfill through heterogeneous compute routing (CPU/GPU), adaptive tuning, UDF-level fault tolerance, and automated storage layout optimization; and (4) hybrid-accelerated data exploration combining a global ID index, transparent OLAP offloading, and background materialized views to deliver second-level point lookups and filtered exports exceeding 20 TB/hour. In production, OmniTable manages over 35 PB of training data across web, code, PDF, and SFT domains, reducing the human-in-the-loop curation cycle from approximately 14 days to approximately 2.5 days (5.6x over the pre-OmniTable production workflow), with consistent feature versioning, auditable lineage, and minimal manual intervention.
\end{abstract}

\maketitle

\pagestyle{\vldbpagestyle}
\begingroup\small\noindent\raggedright\textbf{PVLDB Reference Format:}\\
\vldbauthors. \vldbtitle. PVLDB, \vldbvolume(\vldbissue): \vldbpages, \vldbyear.\\
\href{https://doi.org/\vldbdoi}{doi:\vldbdoi}
\endgroup
\begingroup
\renewcommand\thefootnote{}\footnote{\noindent
This work is licensed under the Creative Commons BY-NC-ND 4.0 International License. Visit \url{https://creativecommons.org/licenses/by-nc-nd/4.0/} to view a copy of this license. For any use beyond those covered by this license, obtain permission by emailing \href{mailto:info@vldb.org}{info@vldb.org}. Copyright is held by the owner/author(s). Publication rights licensed to the VLDB Endowment. \\
\raggedright Proceedings of the VLDB Endowment, Vol. \vldbvolume, No. \vldbissue\ %
ISSN 2150-8097. \\
\href{https://doi.org/\vldbdoi}{doi:\vldbdoi} \\
}\addtocounter{footnote}{-1}\endgroup

\ifdefempty{\vldbavailabilityurl}{}{
\vspace{.3cm}
\begingroup\small\noindent\raggedright\textbf{PVLDB Artifact Availability:}\\
The source code, data, and/or other artifacts have been made available at \url{\vldbavailabilityurl}.
\endgroup
}

\begin{figure*}[t]
  \centering
  \includegraphics[width=\textwidth]{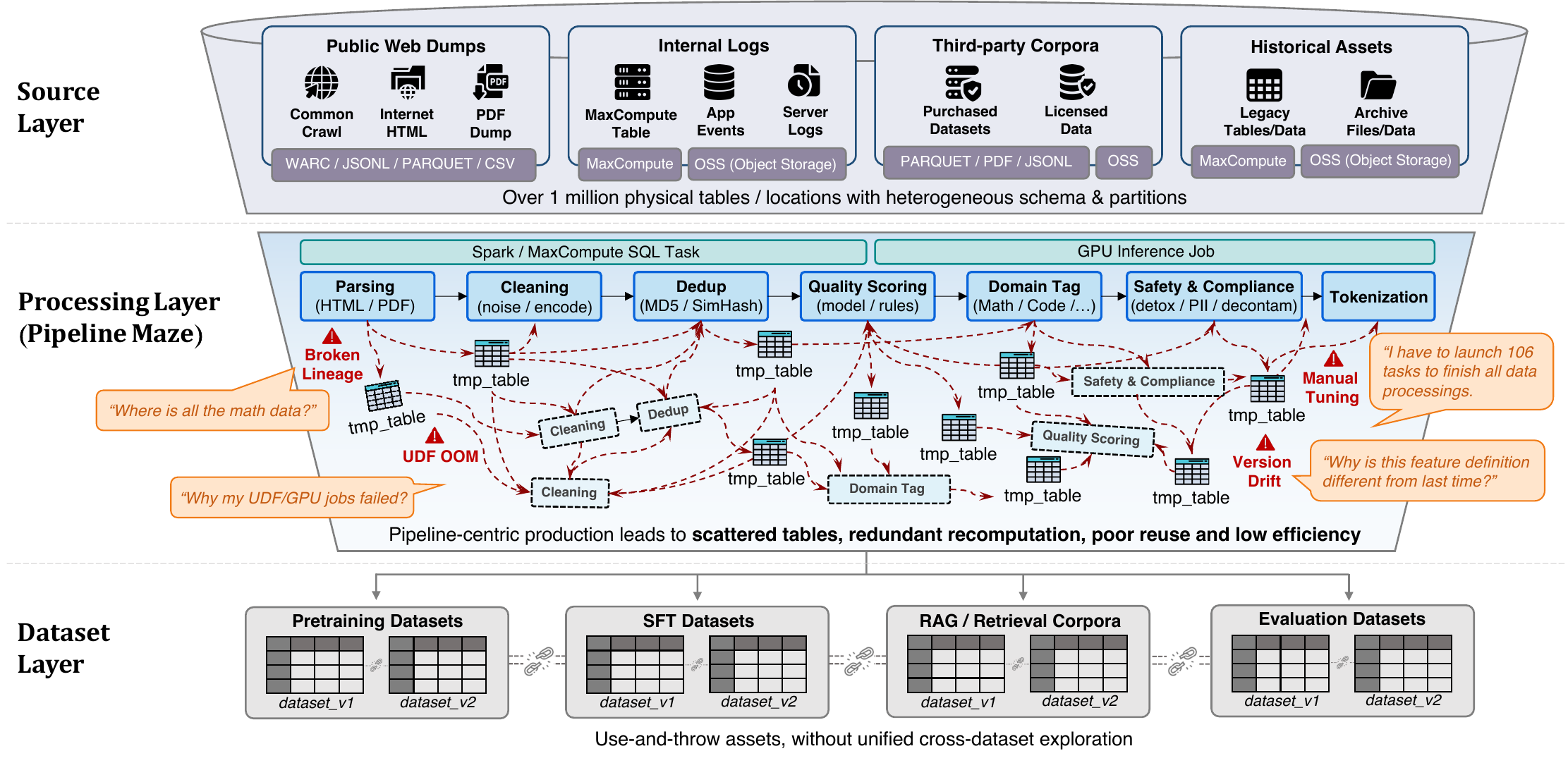}
  \caption{The Pipeline Maze in industrial LLM data preprocessing. Data from heterogeneous sources (top) flows through fragmented processing pipelines (middle), producing isolated, disposable datasets (bottom).}
  \Description{A layered pipeline diagram showing heterogeneous sources flowing through fragmented processing pipelines into isolated datasets.}
  \label{fig:pipeline_maze}
\end{figure*}

\section{Introduction}

Large Language Models (LLMs) such as GPT-4~\cite{Achiam23} and Llama-3~\cite{Dubey24} follow scaling laws~\cite{Kaplan20, Hoffmann22}: their capabilities improve with more parameters, compute, and training data~\cite{Brown20, Muennighoff23}. As architectures converge~\cite{Vaswani17}, \emph{data curation}---collecting, cleaning, enriching, and filtering heterogeneous corpora---has become a key bottleneck~\cite{Penedo24, Liu24, Polyzotis18, Whang23}. At petabyte (PB) scale, structured-data ETL paradigms are inadequate for LLM governance~\cite{Zaharia24}. Existing lakehouse formats (e.g., Iceberg/Delta) provide ACID and schema evolution but are largely \emph{table-centric}. In LLM curation, iteration is \emph{feature}-centric: columns are computed via UDF dependency DAGs and often require CPU/GPU routing. OmniTable targets these gaps with Catalog-driven logical unification and feature-centric execution.

In practice, industrial LLM data preparation workflows often degrade into what we term a \emph{pipeline maze}, characterized by three pain points. \textbf{(1)~Data silos}: corpora from dozens of sources are scattered across hundreds of physical tables, making cross-dataset discovery extremely difficult. \textbf{(2)~Costly feature engineering}: adding a single new feature requires manually coordinating tasks across every relevant dataset---in one case, an engineer had to ``drag and drop 106 tables onto the task canvas'' for a single feature. \textbf{(3)~Broken lineage}: UDF logic is spread across codebases without centralized version control, so feature definitions become inconsistent, and data lineage and feature definitions are not systematically captured; as a result, teams cannot reliably trace how each data/feature iteration affects downstream training runs and model quality~\cite{Sculley15, Schelter17, Vartak16}.

This paper makes the following contributions:

\begin{itemize}
\item \textbf{Unified Wide-Table Abstraction.} A logical data model that consolidates multi-source raw data, multi-stage processing results, and thousands of derived features under a single schema, eliminating data silos (\S\ref{sec:catalog}--\ref{sec:ingestion}).
\item \textbf{Feature Lifecycle Management.} A feature-centric paradigm in which users declare computation logic and the system automatically resolves dependencies, generates execution plans, performs operator fusion, and maintains full lineage (\S\ref{sec:feature_engine}).
\item \textbf{Autonomous Governance and Adaptive Execution Engine.} An engine that intelligently routes tasks across CPU/GPU backends with adaptive tuning and UDF-level fault tolerance, coupled with autonomous background storage optimization for long-term PB-scale scalability (\S\ref{sec:governance}).
\item \textbf{High-Performance Hybrid-Accelerated Exploration.} A query service combining global ID indexing, transparent OLAP offloading, and background materialization to deliver second-level point lookups and filtered exports exceeding 20\,TB/hour (\S\ref{sec:exploration}).
\end{itemize}

The rest of this paper is organized as follows. \S2 motivates the design through real-world challenges. \S3 presents the system overview. \S4 details the design and implementation. \S5 reports experimental results. \S6 shares lessons learned. \S7 discusses related work, and \S8 concludes.

\section{Motivation and Challenges}

\subsection{Workflow of LLM Data Preprocessing}

In industrial LLM development, preprocessing resembles a funnel~\cite{Touvron23, Penedo23, LeScao22}: raw corpora undergo parsing, cleaning, deduplication~\cite{Broder97, Charikar02, Lee22}, scoring, tokenization~\cite{Soldaini24, Lian23, Zheng24}, and sample assembly. In practice it becomes a fragmented \emph{pipeline maze} without a unified abstraction, where each iteration produces disposable tables and ad-hoc jobs.

Figure~\ref{fig:pipeline_maze} summarizes the flow from heterogeneous sources~\cite{Laurencon22, Longpre23, ASF13} through multiple engines (Spark~\cite{Zaharia16}, MaxCompute SQL~\cite{Weng26}, GPU inference) to isolated outputs. This leads to hard discovery, per-table feature backfill (e.g., adding PPL~\cite{Chowdhery23, DeepSeek24}), and difficult fault localization due to missing lineage~\cite{Sculley15, Polyzotis18}.

\subsection{Challenges}

We distill the above pain points into four challenges that define OmniTable's design requirements.

\textbf{C1: Heterogeneity and fragmentation.} LLM corpora come from diverse sources (public crawls, internal logs, procured data) in heterogeneous storage formats and are processed on both CPU and GPU engines with inconsistent runtime environments and failure semantics~\cite{Liu24, Kandel11}. Without a unified data abstraction, each source or engine requires a separate pipeline, preventing users from obtaining a consistent view across sources and processing stages.

\textbf{C2: Scale and performance.} The system must handle hundreds of PB with continuous writes at PB/day scale and thousands of logical columns~\cite{Thusoo10, Dean08}, while simultaneously supporting interactive exploration (second-level point lookups) and high-throughput export (${\geq}$20\,TB/h). Frequent incremental writes introduce small-file and partition explosion, requiring scalable performance under mixed workloads.

\textbf{C3: Agility and iteration speed.} Researchers need to frequently validate different data subsets, filtering strategies, and feature sets through ablation experiments, ideally on a daily basis~\cite{Ratner17}. In traditional workflows, however, the unit of iteration is individual tables and tasks, so adding or modifying a feature requires re-orchestrating pipelines across all affected datasets, with cost scaling linearly or super-linearly with the number of datasets. The system should shift the iteration unit from physical tables to declarative feature operations.

\textbf{C4: Manageability and traceability.} Feature logic is distributed across codebases without centralized version control. Stale lineage~\cite{Vartak16, Schelter17} causes definition drift and irreproducible results. The full feature lifecycle---from declaration through computation to backfill---must automatically generate queryable lineage records for troubleshooting, auditing, and reproduction.

\section{System Overview}

\begin{figure*}[t]
  \centering
  \includegraphics[width=\textwidth]{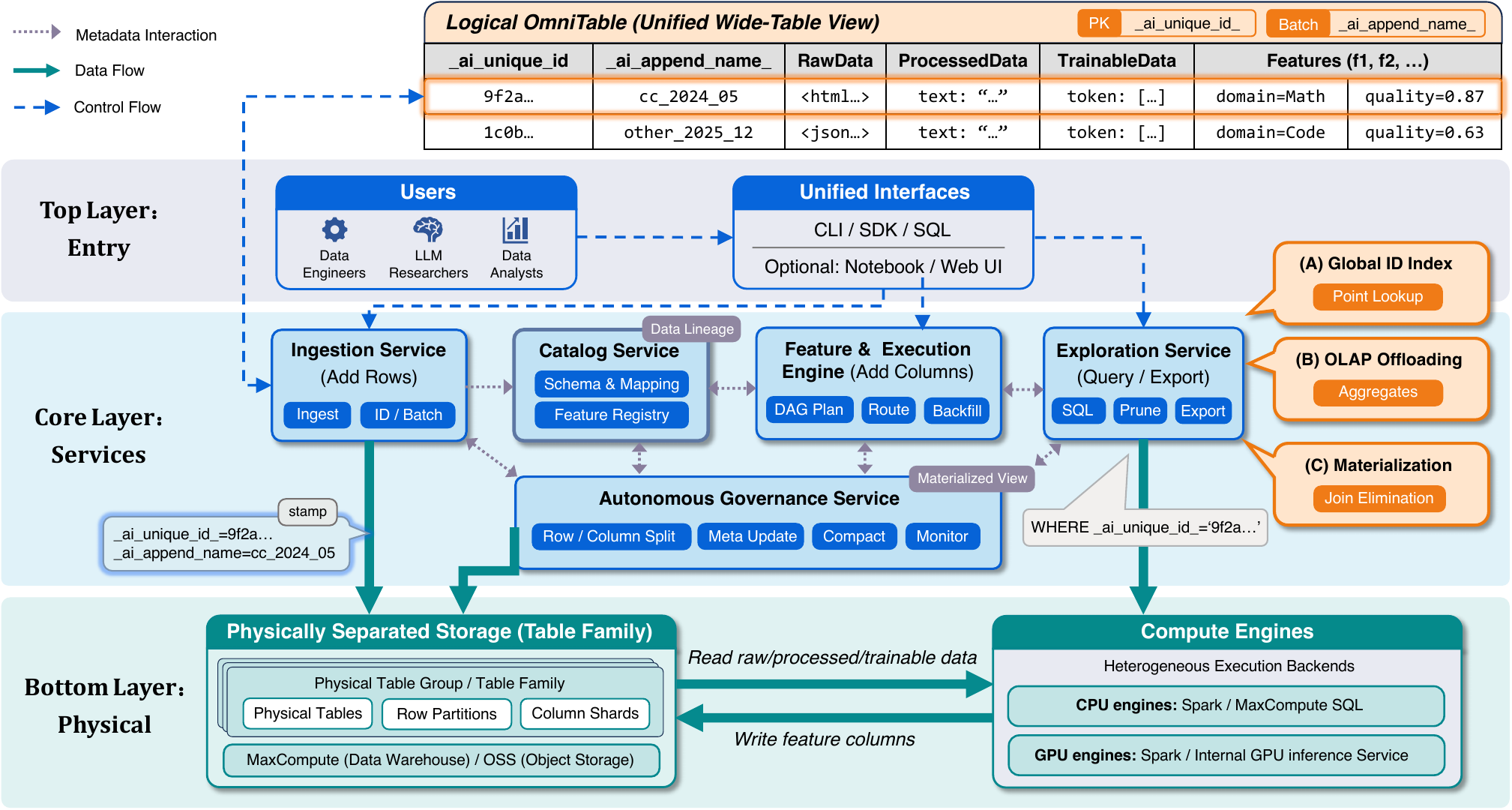}
  \caption{OmniTable system architecture. The Catalog Service acts as the central metadata plane, coordinating Ingestion (add rows), Feature \& Execution Engine (add columns), Exploration Service (query and export), and Autonomous Governance Service (background optimization). The bottom layer shows physically separated storage and heterogeneous compute engines.}
  \Description{Architecture diagram with Catalog coordinating ingestion, feature execution, exploration, governance, storage, and compute engines.}
  \label{fig:architecture}
\end{figure*}

To address the challenges identified in \S2, OmniTable adopts a unified wide table as its core abstraction and, crucially, makes the system \emph{feature-centric}: columns are first-class assets with registered UDF definitions, dependency DAGs, versioning, and engine preferences (CPU/GPU). This shifts the lifecycle from pipeline-centric “tables and tasks” to continuous evolution of data assets (rows) and feature assets (columns).

\subsection{Logical Unification, Physical Separation}

OmniTable presents a single logical wide-table view while allowing diverse physical layouts underneath. Physical separation and query-driven materialization may add ${\sim}$8--15\% storage for hot column groups, but reduce end-to-end curation latency by 5.6$\times$ and cut manual operations by 73\% (\S\ref{sec:exploration}).

\textbf{Logical unification.} All sources, stages, and derived features appear as columns of one logical table. Users operate via \emph{Ingestion} (add rows with batch registration) and \emph{Feature Engineering} (add columns declaratively with automated dependencies and lineage).

\textbf{Physical separation.} The Catalog (\S\ref{sec:catalog}) maps logical columns to physical \emph{table families}, enabling autonomous split/merge/materialization to handle small files, column limits, and mixed workloads without changing logical semantics.

\subsection{Logical Data Model}

OmniTable's logical data model is built around four principles: global primary key alignment, stage-aware data evolution, unbounded feature column expansion, and batch-level governance. Each row represents a traceable data entity; each column represents either the entity's state at a processing stage or a derived feature value. The unified schema is illustrated by the following DDL:

\begin{lstlisting}[style=sqlstyle, caption={OmniTable logical schema (illustrative).}, label={lst:schema}]
CREATE TABLE OmniTable (
  -- Global Primary Key & Lineage Anchor
  _ai_unique_id_   STRING NOT NULL,
  -- Ingestion Batch / Source Tag
  _ai_append_name_ STRING NOT NULL,

  -- Core Data Columns (Stage-aware)
  RawData          STRUCT<...>,
  ProcessedData    STRUCT<...>,
  TrainableData    STRUCT<...>,

  -- Dynamically Growing Feature Columns
  Feature_1        <Type1>,
  ...
  Feature_N        <TypeN>
);
\end{lstlisting}

\texttt{\_ai\_unique\_id\_} serves as the global primary key for alignment across sources, stages, and features, enabling feature backfill joins, point lookups, and full lineage auditing. The three core column groups---\texttt{RawData}, \texttt{ProcessedData}, and \texttt{TrainableData}---represent the original payload, cleaned intermediate representation, and trainable form respectively, with optional version suffixes (e.g., \texttt{ProcessedData\_v2}) to capture processing iterations as versioned columns.

Feature columns \texttt{Feature\_1}\ldots\texttt{Feature\_N} accommodate derived information such as quality scores, domain labels, compliance flags, and deduplication signatures. Explicit columnar representation enables features to be registered, versioned, dependency-resolved, and automatically refreshed, directly serving filtering, statistics, and export.

\texttt{\_ai\_append\_name\_} identifies the batch, source, and version of ingested data, making the batch the basic unit of governance for auditing, task splitting, and query pruning. The DDL above describes a logical contract, not a physical constraint: the Catalog maintains a logical-to-physical mapping, enabling physical splitting, materialization, and indexing transparently (\S\ref{sec:governance}).
\subsection{System Architecture}

As shown in Figure~\ref{fig:architecture}, OmniTable employs a service-oriented architecture centered on the Catalog, consisting of five core services that collaborate to form an end-to-end closed loop of ingestion (add rows), feature backfill (add columns), exploration (query and export), and governance (background optimization).

\textbf{Catalog Service.} The Catalog serves as the authoritative metadata plane and consistency decision point. It maintains the logical wide-table schema, logical-to-physical mappings, feature definitions (including column-level dependency DAGs), and registries for indexes and materialized views. All frontend operations and backend optimizations coordinate through the Catalog, decoupling view stability from physical layout evolvability.

\textbf{Ingestion Service.} This service provides a unified row-addition entry point for multi-source heterogeneous data (MaxCompute tables, OSS~\cite{Li23fast} files). Beyond data transfer, it performs field mapping, generates the global primary key \texttt{\_ai\_unique\_id\_}, tags each record with \texttt{\_ai\_append\_name\_}, and atomically registers batch metadata in the Catalog, transforming discrete datasets into governed asset units.

\textbf{Feature \& Execution Engine.} The engine transforms declarative feature definitions into physical execution plans. It parses dependencies via the Catalog, constructs a minimum-closure execution DAG in topological order, and intelligently routes tasks to heterogeneous backends---Spark/MaxCompute SQL~\cite{Zaharia16, Armbrust15} for CPU-side ETL and UDF computation, or the GPU inference platform for model scoring. During execution, the engine manages concurrency control, checkpoint resumption, failure retries, and result write-back, then synchronizes column status, version, and lineage updates to the Catalog.

\textbf{Exploration Service.} This service handles mixed workloads including interactive exploration, anomaly backtracking, and training data export. It translates SQL on logical wide tables into physical execution plans using Catalog metadata, applying predicate pushdown~\cite{Palkar18}, column pruning~\cite{Abadi08}, and batch pruning. Three complementary acceleration paths are selected automatically: a global ID index for second-level point lookups on \texttt{\_ai\_unique\_id\_}, transparent OLAP offloading (via ClickHouse~\cite{Alexyev24}) for aggregation queries, and background materialized views to eliminate runtime JOINs in filtered exports.

\textbf{Autonomous Governance Service.} This background service maintains long-term performance without user intervention. It monitors storage patterns (small-file density, partition skew, column growth) and query patterns (hot column combinations, frequent filters), then triggers transactional background optimizations---small-file compaction, row/column splitting, and materialized view construction~\cite{Pavlo17}. All changes are isolated from frontend operations through Catalog versioning and registered back upon completion, allowing subsequent queries to transparently benefit.

\section{System Design and Implementation}
\label{sec:design}

This section details the internal design and key implementation of each core service, organized by the data lifecycle: Catalog (\S\ref{sec:catalog}), Ingestion (\S\ref{sec:ingestion}), Feature \& Execution Engine (\S\ref{sec:feature_engine}), Exploration (\S\ref{sec:exploration}), and Autonomous Governance (\S\ref{sec:governance}).

\subsection{Catalog Service}
\label{sec:catalog}

The Catalog Service is the authoritative metadata plane for OmniTable. All frontend operations (ingestion, feature backfill, query export) and backend optimizations (splitting, merging, materialization) use the Catalog as a unified coordination point: frontend services obtain logical-to-physical mappings to generate correct execution plans, while backend services register optimization results so that frontend queries transparently benefit. This design decouples logical view stability from physical layout evolvability.

\subsubsection{Logical-Physical Schema Mapping}

Users see a single logical wide table with a unified schema, but this table is physically backed by multiple tables---a \emph{Table Family}. The Catalog maintains a five-layer entity model spanning LogicalTable, LogicalColumn, PhysicalTableGroup, PhysicalTable, and PhysicalColumn. A LogicalTable corresponds to a complete user-facing wide table (e.g., \texttt{aidata:// \allowbreak tables/web}). Each LogicalColumn carries semantic information (name, type, analysis type, comments). A PhysicalTableGroup is an intermediate layer grouping physical tables by function or batch. Each PhysicalTable resides on the underlying engine (MaxCompute or OSS) with attributes such as storage path, partition information, and file format. PhysicalColumns are linked to LogicalColumns through bidirectional references in the Catalog.

The key advantage of this multi-layer mapping is physical layout evolvability. When the backend governance service (\S\ref{sec:governance}) detects that a physical table exceeds the storage engine's column limit (e.g., MaxCompute's ${\sim}$1200-column limit), the system automatically performs column splitting by updating only the affected mapping pointers in the Catalog, without modifying any logical schema. Similarly, small-partition merges atomically update PhysicalTable entries after data rewrite. The query optimizer always obtains the latest mapping to generate plans, making physical changes fully transparent. Each mapping entry carries a version number and timestamp for historical backtracking. The Catalog stores metadata in a relational database with optimistic locking for concurrent update consistency; metadata query latency remains at the millisecond level in production.

The Catalog also manages batch (Append) metadata. Each ingestion registers an Append entry recording the batch name (\texttt{\seqsplit{\_ai\_append\_name\_}}), associated PhysicalTable identifiers, row count, ingestion time, and source information. This makes \texttt{\_ai\_append\_name\_} a key dimension for batch pruning, incremental backfill scheduling, and audit backtracking: queries filtering on a specific batch (e.g., \texttt{WHERE \_ai\_append\_name\_ = 'cc'}) can directly locate the corresponding physical table subset, significantly reducing I/O.

\begin{lstlisting}[style=sqlstyle, float=tbp, basicstyle=\ttfamily\footnotesize, caption={Feature definition schema (simplified).}, label={lst:featdef},
  morekeywords={inputColumns, outputColumns, name, type, analysisType, comment, defaultFeatExpression, computeEngine, tablePath, extraInfo, hints, parallel, where_condition, FeatureDefinition},
  keywordstyle=\color{sqlkeyword}\bfseries]
FeatureDefinition {
  inputColumns:  [col_1, col_2, ...],
  outputColumns: [{
    name: <column_name>, type: <data_type>,
    analysisType: <semantic_type>,
    comment: <description>
  }, ...],
  defaultFeatExpression: <expr>,
  computeEngine: <engine_preference>,
  tablePath: <wide_table_path>,
  extraInfo: { hints: <config>,
    parallel: <N>, where_condition: <filter> }
}
\end{lstlisting}

\subsubsection{Feature Metadata and Lineage}

OmniTable treats features as first-class metadata~\cite{Helland15}. Complete feature \seqsplit{definitions---computation} logic, input/output dependencies, runtime preferences, and version history---are atomically registered in the Catalog. This centralizes feature computation logic that is traditionally scattered across codebases and enables automated lineage tracing and dependency resolution.

A feature is registered via the \texttt{omni-cli add feature} command with a structured definition specifying input columns, output columns (with name, type, and semantic annotation), computation expression, engine preference, and optional runtime hints (Listing~\ref{lst:featdef}).

\begin{figure}[t]
  \centering
  \includegraphics[width=0.68\columnwidth]{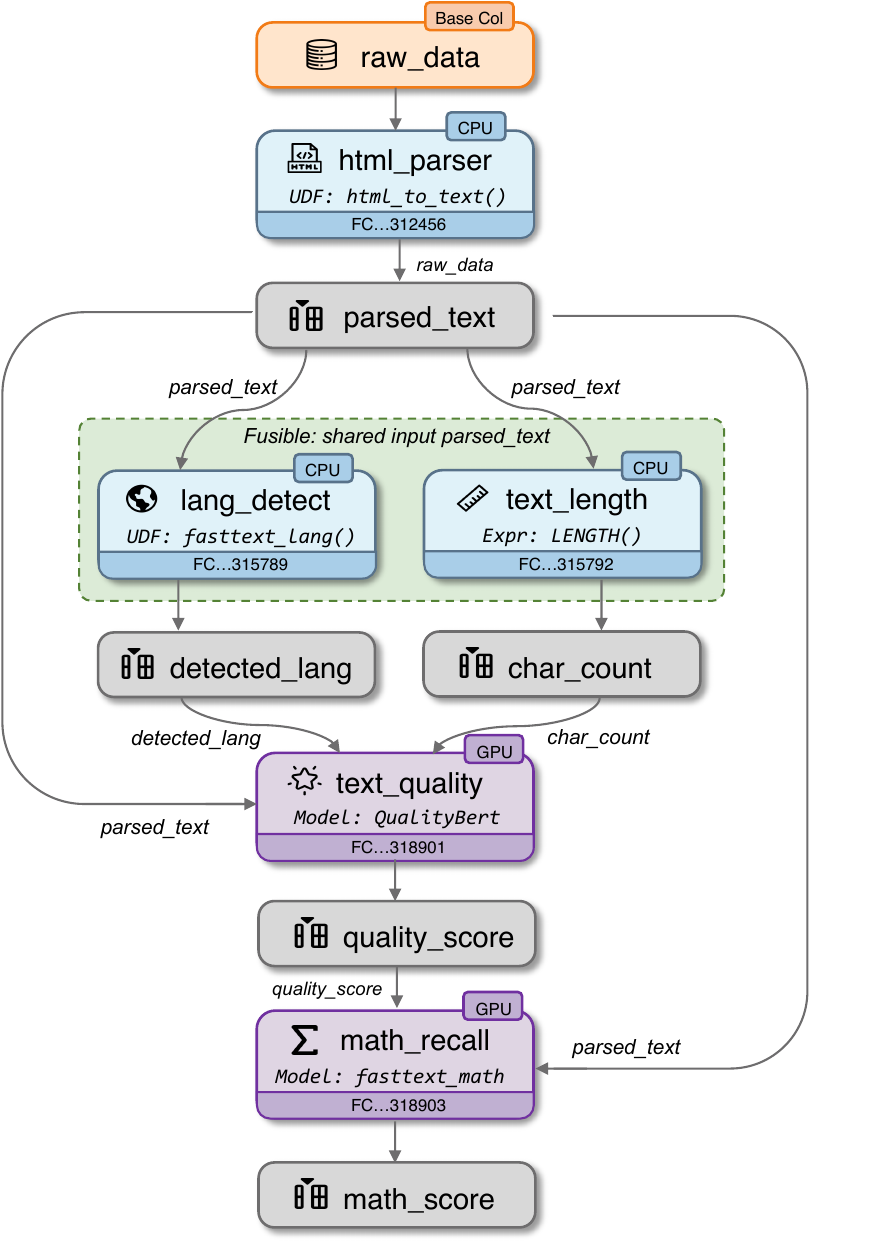}
  \caption{Feature computation DAG automatically constructed by the Catalog, showing column-level dependencies among registered features on the web wide table.}
  \Description{Directed acyclic graph showing dependencies from raw data through parsed text to language, length, quality, and math-recall features.}
  \label{fig:feature_dag}
\end{figure}

Based on registered definitions, the Catalog automatically constructs a column-level dependency DAG. Figure~\ref{fig:feature_dag} illustrates a real subgraph from the web wide table: \texttt{\seqsplit{raw\_data}} $\rightarrow$ \texttt{\seqsplit{html\_parser}} $\rightarrow$ \texttt{\seqsplit{parsed\_text}}, which fans out to \texttt{\seqsplit{lang\_detect}} and \texttt{\seqsplit{text\_length}}; \texttt{\seqsplit{quality\_score}} depends on both \texttt{\seqsplit{parsed\_text}} and \texttt{\seqsplit{detected\_lang}}; and \texttt{\seqsplit{math\_recall}} depends on \texttt{\seqsplit{parsed\_text}} and \texttt{\seqsplit{quality\_score}}. The DAG is stored as an adjacency list, with each node recording its feature ID, column ID, and parent column IDs.

\subsection{Ingestion Service}
\label{sec:ingestion}

The Ingestion Service integrates heterogeneous corpora into the logical wide table through a declarative interface (\texttt{omni-cli submit}). Users specify source type, location, target wide table, and column mapping (e.g., \texttt{body\_text:raw\_data}). Unmapped source columns are ignored; missing logical columns are filled with \texttt{NULL} for subsequent backfill. Internally, a three-stage pipeline handles source adaptation (format parsing and validation), schema normalization, and physical writing (columnar organization for bulk loads, fast append for incremental writes). New wide tables are created on demand.

\texttt{\_ai\_unique\_id\_} serves as the global primary key. The default strategy uses deterministic content hashing (MD5 of \texttt{raw\_data}), providing natural deduplication without coordination. A random UUID strategy is available when content deduplication is unnecessary (e.g., SFT data). Ingestion uses write-then-commit semantics: parallel shards process data, then the service atomically registers an Append entry in the Catalog recording batch name, physical tables, row count, source, and mapping. Data becomes visible only after commit; failures trigger rollback. \texttt{\_ai\_append\_name\_} serves as scheduling granularity for backfill (\S\ref{sec:feature_engine}), query pruning (\S\ref{sec:exploration}), and governance (\S\ref{sec:governance}). In production, the web wide table has ingested over 25\,PB across 200+ batches.

\subsection{Feature and Execution Engine}
\label{sec:feature_engine}

The Feature \& Execution Engine shifts feature engineering from manual, table-by-table pipeline orchestration to declarative operations on logical wide tables. Users declare the computation logic of a feature; the system automatically resolves dependencies, generates execution plans, routes tasks to heterogeneous backends, and manages fault tolerance.

\begin{algorithm}[!t]
\small
\setlength{\algomargin}{0.8em}
\SetAlgoInsideSkip{smallskip}
\caption{Declarative Feature Backfill}
\label{alg:backfill}
\KwIn{BackfillPlan $P = \{\textit{tablePath}, \{(a_i, c_j)\}\}$; Feature DAG $G$ from Catalog}
\KwOut{Computed feature columns; updated Catalog metadata}

\tcc{Stage 1: Dependency Resolution}
$Q \leftarrow$ copy of $P.\textit{pairs}$\;
\ForEach{$(a_i, c_j) \in Q$}{
    $\textit{Ancestors} \leftarrow G.\text{transitiveClosure}(c_j)$\;
    \ForEach{$f_k \in \textit{Ancestors}$}{
        \If{$\text{Catalog.status}(a_i, f_k) \neq \textsc{Computed}$}{
            $Q \leftarrow Q \cup \{(a_i, f_k)\}$\;
        }
    }
}
$D \leftarrow \text{topologicalSort}(Q, G)$\;

\tcc{Stage 2: Physical Plan Generation}
\ForEach{$(a_i, c_j) \in D$ in topological order}{
    $(P_{\textit{in}}, P_{\textit{out}}) \leftarrow \text{Catalog.resolvePhysical}(a_i, c_j)$\;
    $\textit{engine} \leftarrow \text{RouteEngine}(c_j.\textit{computeEngine},\; c_j.\textit{mode})$\;
    $\textit{params} \leftarrow \text{AdaptiveTune}(c_j,\; a_i,\; \textit{engine})$\;
    $T_{ij} \leftarrow \text{buildTask}(P_{\textit{in}}, P_{\textit{out}}, c_j.\textit{expr}, \textit{engine}, \textit{params})$\;
    \If{$\exists\; (a_i, c_m) \in D$ sharing $P_{\textit{in}}$ and engine}{
        $T_{ij} \leftarrow \text{fuseOperators}(T_{ij},\; T_{im})$\;
    }
}

\tcc{Stage 3: Scheduled Execution}
\While{$\exists$ unfinished node in $D$}{
    $T_{ij} \leftarrow$ next ready task\;
    Dispatch $T_{ij}$ with UDF-level fault tolerance\;
    \eIf{$T_{ij}$ succeeds}{
        Write results; record checkpoint\;
    }{
        \lIf{retries $<$ max}{Re-enqueue}
        \lElse{Mark \textsc{Failed}; log to ErrorTable}
    }
}

\tcc{Stage 4: Metadata Commit}
\ForEach{completed $T_{ij}$}{
    Catalog.atomicUpdate$(a_i, c_j)$: status, physicalMapping, lineage\;
}
\end{algorithm}

\subsubsection{Declarative Feature Lifecycle}

The user interaction reduces to two operations: registering a feature definition via \texttt{omni-cli add feature} (\S\ref{sec:catalog}) and submitting a backfill plan via \texttt{omni-cli add jobs}. The system then executes the four-stage process shown in Algorithm~\ref{alg:backfill}.

\textbf{Stage 1: Dependency Resolution.} The user's backfill plan typically specifies only the target feature and batch (e.g., compute \texttt{math\_recall} on batch \texttt{cc}). The system traverses the feature DAG to collect the transitive closure, checks each predecessor's computation status, and auto-expands the plan to include all uncomputed dependencies. Formally, given a user-specified set of backfill targets $P \subseteq \mathcal{A} \times \mathcal{C}$ (append--column pairs) and the feature dependency DAG $G = (\mathcal{C}, E)$, the minimum-closure backfill set is:

\begin{equation}
\label{eq:closure}
\begin{split}
Q^{*} = \bigl\{(a_i, c_k) \;\big|\; & (a_i, c_j) \in P,\;\; c_k \in \mathrm{Anc}_G(c_j) \cup \{c_j\},\\
& \mathrm{status}(a_i, c_k) \neq \textsc{Computed}\bigr\}
\end{split}
\end{equation}

where $\mathrm{Anc}_G(c_j)$ denotes the set of all ancestor nodes of $c_j$ in $G$. Topological sorting of $Q^{*}$ then produces an execution DAG respecting dependency order.

\textbf{Stage 2: Physical Plan Generation.} For each node in the DAG, the system resolves input/output physical tables via the Catalog, selects the execution engine based on the feature's \texttt{computeEngine} and operator characteristics, and sets resource parameters via adaptive tuning (\S\ref{sec:tuning}). An important optimization is \emph{operator fusion}: when multiple features in the same batch share input columns and target engine (e.g., \texttt{lang\_detect} and \texttt{text\_length} both reading \texttt{parsed\_text} on Spark), the system merges them into a single SQL task computing multiple output columns in one scan, reducing redundant I/O by several times in production.

\textbf{Stage 3: Scheduled Execution.} The scheduler dispatches ready tasks (all parents completed) according to topological order and user-specified concurrency. Each task runs with UDF-level fault tolerance (\S\ref{sec:fault_tolerance}). On success, results are written and partition-level checkpoints are recorded, supporting resumption after interruption. Failed tasks are retried up to a configured limit; beyond that, they are marked failed with details logged to an error table.

\textbf{Stage 4: Metadata Commit.} Upon completion, the system atomically updates the Catalog for each \texttt{(append, column)} pair: setting status to \textsc{Computed}, registering the output physical mapping, and recording lineage (feature ID, version, timestamp, engine). This transforms computation results from temporary output into queryable, traceable assets.

\subsubsection{Intelligent Routing for Heterogeneous Compute}
\label{sec:routing}

Feature computation spans rule-based statistics (CPU-suited), deep learning scoring (GPU-required), and Python UDFs with complex dependencies (e.g., fastText~\cite{Joulin17}, BERT~\cite{Devlin19}). Users declare \texttt{\seqsplit{-{}-mode=cpu\_only|gpu\_only|all}} during registration; the system routes at runtime based on operator profile (UDF dependency analysis), engine capabilities (Spark/MaxCompute SQL~\cite{Armbrust15} vs.\ GPU inference), and cluster load~\cite{Binnig15}. In production, ${\sim}$70\% of features route to CPU clusters, ${\sim}$20\% (fastText/BERT) are placed dynamically based on model size and resource conditions, and ${\sim}$10\% (e.g., Qwen~\cite{Qwen25} PPL) route to GPU. Users remain unaware of engine selection.

\subsubsection{Elasticity and Fault Tolerance}
\label{sec:fault_tolerance}
\label{sec:tuning}

PB-scale feature backfill faces two systemic risks: improper resource configuration causing OOM or resource waste, and bad data (e.g., extremely long texts, encoding errors) causing entire batch tasks to fail. OmniTable addresses both through \emph{adaptive parameter tuning} and \emph{UDF-level fault tolerance}.

\textbf{Adaptive Parameter Tuning.} The engine computes resource parameters from two signals. First, heuristic rules use Catalog statistics (partition count, row count, data volume, average record size) and feature profiles: Python model loading raises Spark's \texttt{\seqsplit{memoryOverhead}}, while high partition counts adjust \texttt{\seqsplit{mapper.split.size}}. Second, History-Based Optimization (HBO)~\cite{Shankhdhar24} records CPU time, peak memory, I/O throughput, feature identifiers, and batch sizes, then interpolates from similar historical runs. This strategy gives reasonable cold-start parameters and improves with history; in production it raised novice users' first-submission success rate from about 60\% to over 90\%.

\textbf{UDF-Level Fault Tolerance.} PB-scale unstructured data inevitably contains bad records, and one such record can fail a Spark task containing millions of inputs. OmniTable wraps each UDF invocation with timeout and memory checks. On Python OOM, timeout, or uncaught exception, it logs the record's \texttt{\seqsplit{\_ai\_unique\_id\_}}, exception type, and stack summary to an error table, writes NULL, and continues. The Catalog records success/failure counts and error-log references after completion, while configurable retries handle transient failures. Adaptive tuning reduces configuration failures before submission; UDF-level isolation contains data failures during execution.

\subsection{Exploration Service}
\label{sec:exploration}

The Exploration Service bridges the gap between the logically unified wide-table view and physically distributed storage, handling mixed workloads: second-level point lookups for anomaly tracing, aggregation queries for feature distribution analysis, and high-throughput filtered exports (${\geq}$20\,TB/h) for ablation experiments. It translates user SQL on logical wide tables into physical execution plans and selects among three complementary acceleration paths.

\subsubsection{Logical Query Translation and Rewriting}

The query optimizer translates user SQL on logical wide tables into plans executable on underlying physical table families. Because logical columns may span multiple physical tables (due to column splitting or batch-based organization), translation involves multi-step reasoning beyond simple table name replacement.

The optimizer parses user SQL to extract referenced columns and predicates, then queries the Catalog for each column's physical location. The result is typically a multi-table JOIN on \texttt{\_ai\_unique\_id\_} with UNION ALL across batches. The optimizer then applies standard rewrite rules: \emph{predicate pushdown}~\cite{Palkar18} pushes filters into physical table scans; \emph{column pruning}~\cite{Abadi08, Stonebraker05} reads only required columns, avoiding I/O on large unused columns such as \texttt{raw\_data}; and \emph{batch pruning} uses Catalog Append metadata to skip physical tables of unrelated batches when \texttt{\_ai\_append\_name\_} filters are present.

After these optimizations, the optimizer attempts to match one of three acceleration paths: global ID index for equality lookups on \texttt{\_ai\_unique\_id\_} (\S\ref{sec:id_index}), OLAP offloading for aggregate queries (\S\ref{sec:olap}), or materialized view matching for multi-column filtered exports (\S\ref{sec:mv}).

\subsubsection{Global ID Index}
\label{sec:id_index}

Sample tracing retrieves a complete record by \texttt{\seqsplit{\_ai\_unique\_id\_}}, including all stage data and feature values. Without an index, it requires full scans across physical tables and takes minutes at PB scale.

OmniTable builds a global secondary index on \texttt{\_ai\_unique\_id\_} using a distributed key-value store (HBase~\cite{ASF08, Chang08}). Each entry maps an ID to its physical table identifier, partition path, and row group offset. Index construction is asynchronous: incremental updates are triggered after successful ingestion or feature backfill commits, with version and coverage information maintained in the Catalog.

When the optimizer detects an equality filter on \texttt{\seqsplit{\_ai\_unique\_id\_}}, it routes to the index path. The index service returns physical locations with millisecond latency; the system reads only target row groups and assembles columns from relevant tables in parallel. In production, the index covers over 300 billion records, with single-ID latency of 5--15 seconds.

\subsubsection{Analytical Query Offloading to OLAP}
\label{sec:olap}

Analytical queries with aggregations are common during filtering strategy design but slow on batch engines (minutes to hours). OmniTable maintains a ClickHouse~\cite{Alexyev24} instance, incrementally synchronizing frequently accessed feature columns based on query pattern analysis. Offloading is transparent: when the optimizer verifies all referenced columns are synchronized with sufficient freshness, it rewrites and routes to ClickHouse; otherwise it falls back to the batch engine. In production, this reduces aggregation response times from minutes to seconds (${\sim}$100$\times$).

\subsubsection{Join Elimination via Background Materialization}
\label{sec:mv}

When query columns span three or more physical tables, runtime JOINs become the dominant bottleneck---each JOIN involves large-scale scans and expensive shuffles. For filtered export queries (typically involving 10+ feature columns), multi-table JOIN overhead reduces throughput well below the 20\,TB/h target.

OmniTable eliminates runtime JOINs through background materialization~\cite{Gupta95, Goldstein01, Halevy01}. The Autonomous Governance Service records column co-occurrence in query logs and periodically constructs a co-occurrence frequency matrix~\cite{Quinlan86}. When a column combination exceeds a frequency threshold and spans multiple physical tables, it becomes a materialization candidate. The service pre-joins these columns on \texttt{\_ai\_unique\_id\_} into a materialized wide table and atomically registers its metadata (covered columns, batch range, version) in the Catalog.

At query time, the optimizer checks whether a materialized view fully covers the required columns. If so, the query is rewritten as a single-table scan, eliminating all JOINs. For partial coverage, the optimizer evaluates whether using the view reduces JOINs (e.g., from three to one) and selects a mixed path accordingly.

Materialized views are maintained via incremental refresh triggered by Catalog change events; delayed refreshes are handled by compensation queries on uncovered increments merged via UNION ALL. In production, ${\sim}$80\% of filtered export queries reference 15--20 feature columns across 3--5 physical tables; after materialization, these reduce to single-table scans with 3--5$\times$ latency improvement and throughput above 20\,TB/h.

\begin{figure}[t]
  \centering
  \includegraphics[width=\columnwidth]{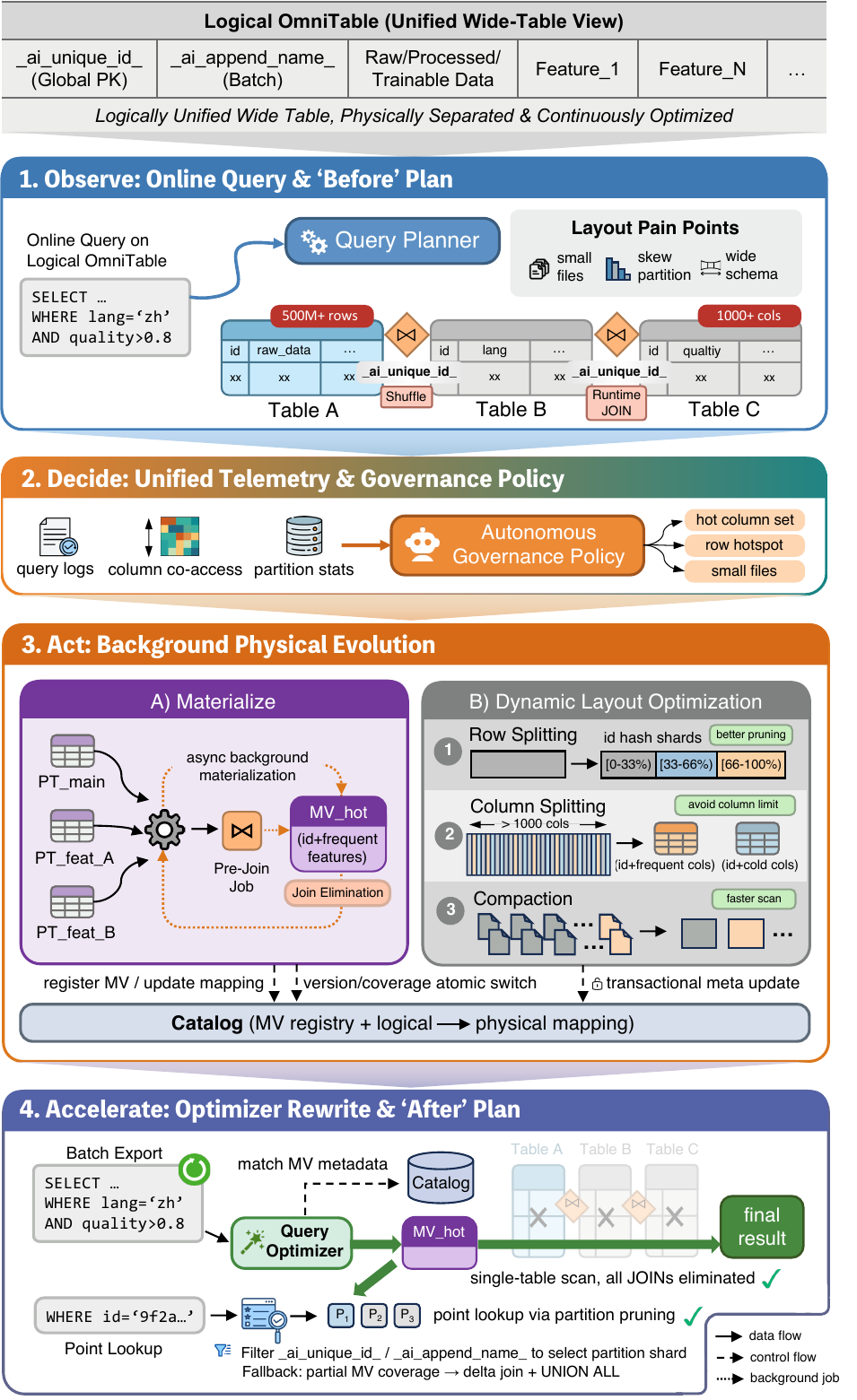}
  \caption{Autonomous governance pipeline: observe layout pain points, decide optimization policy, perform background physical evolution (materialization, splitting, compaction), and accelerate queries via optimizer rewriting.}
  \Description{Pipeline showing observation, optimization policy selection, physical layout evolution, and query acceleration.}
  \label{fig:layout_opt}
\end{figure}

\subsection{Autonomous Governance Service}
\label{sec:governance}

Without continuous maintenance, PB-scale continuous writes and growing column counts inevitably degrade performance: small files reduce scan efficiency~\cite{Ghemawat03}, column counts hit engine limits~\cite{Stonebraker05}, and partition skew causes long-tail queries. The Autonomous Governance Service addresses these challenges through automated background optimization, isolated from frontend operations.

\subsubsection{Transactional Consistency Guarantees}

All backend optimizations follow a three-phase Prepare-Execute-Commit protocol~\cite{Bernstein81}. In the \emph{Prepare} phase, the service acquires an exclusive lock at PhysicalTableGroup granularity from the Catalog; if the target is occupied by a frontend write, the optimization is postponed. Frontend reads do not hold exclusive locks but obtain a consistent metadata snapshot at query start, ensuring in-flight queries are unaffected by concurrent backend changes~\cite{Agrawal87}.

In the \emph{Execute} phase, physical reorganization (merging, splitting, materialization) writes new files to a staging area without modifying existing files. In the \emph{Commit} phase, the service atomically updates the Catalog: registering new physical files, updating logical-to-physical mappings, and marking old files for reclamation. The new layout becomes visible only after successful commit; subsequent queries automatically use the optimized layout. Failures trigger metadata rollback and staging cleanup.

\subsubsection{Dynamic Storage Layout Optimization}

The physical layout continuously adapts to data growth, feature evolution, and query pattern changes through three mechanisms (Figure~\ref{fig:layout_opt}).

\textbf{Row Splitting.} When a physical partition exceeds thresholds (e.g., 500M rows or 500\,GB), the system divides it into $N$ sub-partitions based on \texttt{\_ai\_unique\_id\_} hash ranges. The Catalog records each sub-partition's range, enabling precise partition pruning for ID-based queries and improved parallelism for full scans.

\textbf{Column Splitting.} When logical column count approaches the storage engine's limit (e.g., MaxCompute's ${\sim}$1200 columns), the governance service groups columns by access frequency from query logs. Low-frequency feature columns are migrated to auxiliary physical tables (Column Shards) containing only \texttt{\_ai\_unique\_id\_} and the migrated columns. The main table retains high-frequency and core columns. The Catalog updates logical-to-physical mappings; queries involving only high-frequency columns require no JOINs. In production, the web wide table's 800+ logical columns are distributed across 4--6 physical tables, each with 200--300 columns.

\textbf{Small-File Compaction.} Append-only writes from ingestion and backfill accumulate small files. A background compaction service periodically scans file size distributions and triggers merging when thresholds are exceeded (e.g., $>$1000 files per partition or $>$50\% files smaller than 64\,MB). The compaction task reads small files, reorders data by \texttt{\_ai\_unique\_id\_}, and writes optimally sized files (256\,MB--1\,GB) with columnar encoding and compression~\cite{Elgohary16}. OmniTable uses a \emph{tiered compaction strategy}~\cite{ONeil96}: daily lightweight merges for hot partitions and weekly deep merges for cold partitions, balancing merge benefits against resource costs (approximately 5--8\% of cluster resources in production).

\section{Experimental Evaluation}

\begin{figure*}[t]
  \centering
  \includegraphics[width=\textwidth]{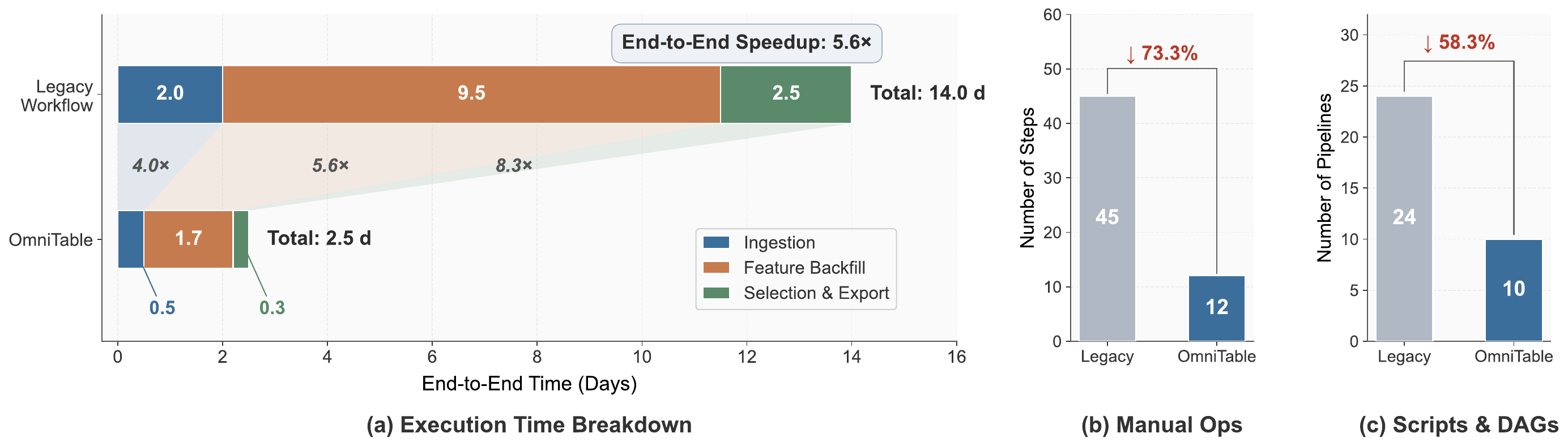}
  \caption{End-to-end SFT data curation comparison. (a)~Execution time breakdown by stage: OmniTable achieves 5.6$\times$ speedup (14.0\,d $\rightarrow$ 2.5\,d). (b)~Manual operation steps reduced by 73\%. (c)~Independent pipelines/scripts reduced by 58\%.}
  \Description{Three charts comparing legacy workflow and OmniTable on execution time, manual steps, and pipeline counts.}
  \label{fig:e2e}
\end{figure*}

\subsection{Experimental Setup}

\textbf{Datasets and Production Environment.} All experiments were conducted on a real production deployment of OmniTable. Table~\ref{tab:datasets} summarizes the core data assets. The web wide table (\texttt{web}) is the largest, covering 13 sources (Common Crawl, internal crawls, licensed third-party corpora, etc.) with over 25\,PB, 300B+ records, 800+ logical columns (200+ registered features), 6 physical tables, and 200+ ingested batches. The code (\texttt{code}), PDF (\texttt{pdf}), and SFT (\texttt{post\_sft}) wide tables cover the remaining major data types, managing a combined total of over 35\,PB.

\textbf{Workload Characteristics.} We evaluate three groups: (1)~governance batch tasks (ingestion, feature backfill, failure recovery); (2)~exploratory queries (point lookups, filtered exports, aggregation statistics); and (3)~scalability stress tests. The 200+ features on \texttt{web} span ${\sim}$70\% lightweight SQL/rule UDFs suited for CPU/Spark~\cite{Zaharia12, Zaharia16}, ${\sim}$20\% CPU-based model inference UDFs (fastText~\cite{Joulin17}, BERT~\cite{Devlin19}), and ${\sim}$10\% GPU-intensive operators (Qwen~\cite{Qwen25} PPL).

\textbf{Hardware.} CPU side: MaxCompute/Spark hybrid cluster with ${\sim}$12,000 nodes (64-core CPU, 256\,GB RAM each). GPU side: ${\sim}$300 NVIDIA L20 cards (48\,GB VRAM). Storage: MaxCompute warehouse and OSS object storage using Parquet~\cite{ASF13} columnar format. Catalog: high-availability MySQL cluster. ID index: HBase~\cite{ASF08} cluster. OLAP: ClickHouse~\cite{Alexyev24} cluster. Spark task parameters are set automatically by adaptive tuning unless noted otherwise.

\textbf{Baseline (Legacy Workflow).} We defined a baseline replicating the pre-OmniTable production workflow (\S2): (1)~data organized in independent MaxCompute tables without unified schema; (2)~feature engineering via manual pipeline orchestration on a visual canvas; (3)~cross-dataset queries via manually written multi-table JOINs; (4)~task tuning via engineer experience. Efficiency data is derived from historical task records and operation logs; one manual retry is allowed per failure for fairness. This comparison targets \emph{workflow-level coordination cost}---the overhead of locating tables, wiring pipelines, diagnosing failures, and re-submitting jobs---which is orthogonal to compute-level optimizations in systems such as Data-Juicer~\cite{Chen24} or Delta Lake~\cite{Armbrust20}. Per-subsystem performance is validated separately in \S\ref{sec:scalability}--\ref{sec:exploration_eval}.

\textbf{Ablation (OmniTable-NoGov).} To isolate the contribution of autonomous governance, we disable background compaction, row splitting, and column splitting, retaining only logical query translation and basic pruning. This quantifies the necessity of physical layout evolution at PB scale (\S\ref{sec:scalability}).

\textbf{Methodology.} All time results are medians of three runs. Exploration queries are replayed at fixed concurrency (20) within the same time window; error bars show cross-run variation.

\begin{table}[t]
  \centering
  \caption{Core data assets used in the experiments.}
  \label{tab:datasets}
  \small
  \begin{tabular}{lrrrrc}
    \toprule
    \textbf{Table} & \textbf{Records} & \textbf{Size} & \textbf{Log.\ Cols} & \textbf{Phys.\ Tbl.} & \textbf{Batches} \\
    \midrule
    \texttt{web}      & 300B+  & 25\,PB  & 800+ & 6 & 200+ \\
    \texttt{code}          & 3.4B   & 3.8\,PB & 350  & 4 & 85   \\
    \texttt{pdf}           & 1.8B   & 5.2\,PB & 280  & 3 & 62   \\
    \texttt{post\_sft}     & 210M   & 0.8\,PB & 120  & 3 & 45   \\
    \midrule
    \textbf{Total}         & \textbf{305B+} & \textbf{35\,PB} & --- & \textbf{16} & \textbf{392+} \\
    \bottomrule
  \end{tabular}
\end{table}

\subsection{End-to-End Data Curation Efficiency}
\label{sec:e2e}

We compare OmniTable and Legacy Workflow on a real SFT data processing scenario: collecting instruction data from 8 sources, computing 12 features (9 CPU UDFs, 3 GPU inference) covering quality scoring, safety compliance, and domain classification, and exporting a high-quality subset for ablation experiments. This scenario covers all three stages---ingestion, feature backfill, and filtered export. Input batches and filtering thresholds are fixed to ensure identical business objectives.

\textbf{Legacy Workflow.} Ingestion takes ${\sim}$2 days to locate 8 source tables, write conversion scripts, and copy data. Feature backfill is the main bottleneck at ${\sim}$9.5 days: engineers configure 12 feature nodes $\times$ 8 tables (${\sim}$96 nodes) on the orchestration canvas, repeatedly handling failures from improper parameters or bad data. Without UDF-level fault tolerance, a single anomalous record fails the entire TB-scale batch, triggering costly investigate-retry cycles. Filtered export takes ${\sim}$2.5 days writing manual multi-way JOINs across 8 result tables. Total: ${\sim}$14 days, ${\sim}$45 manual steps, 24 pipelines, 35 physical tables.

\textbf{OmniTable.} Ingestion takes ${\sim}$0.5 days via 8 \texttt{omni-cli submit} commands. Feature backfill takes ${\sim}$1.7 days: after registering 12 feature definitions, a single backfill plan triggers automatic dependency resolution, heterogeneous engine routing, adaptive tuning, and UDF-level fault tolerance---no manual pipeline orchestration required. Operator fusion merges features sharing input columns into single SQL tasks. Filtered export takes ${\sim}$0.3 days via a single SQL query on the logical wide table with automatic materialized view matching. Total: ${\sim}$2.5 days, ${\sim}$12 manual steps, 10 commands, 1 logical wide table.

\textbf{Results.} Figure~\ref{fig:e2e} shows that OmniTable achieves a 5.6$\times$ end-to-end speedup over the Legacy Workflow. The feature backfill stage contributes the largest gain (9.5\,d $\rightarrow$ 1.7\,d, 5.6$\times$), driven by declarative lifecycle management eliminating per-table orchestration and adaptive fault tolerance eliminating manual failure recovery. Manual steps decrease by 73.3\% (45 $\rightarrow$ 12), pipelines by 58.3\% (24 $\rightarrow$ 10), and engineers are freed from overnight pipeline monitoring.

\subsection{PB-Scale Performance and Scalability}
\label{sec:scalability}

\begin{figure*}[t]
  \centering
  \includegraphics[width=\textwidth]{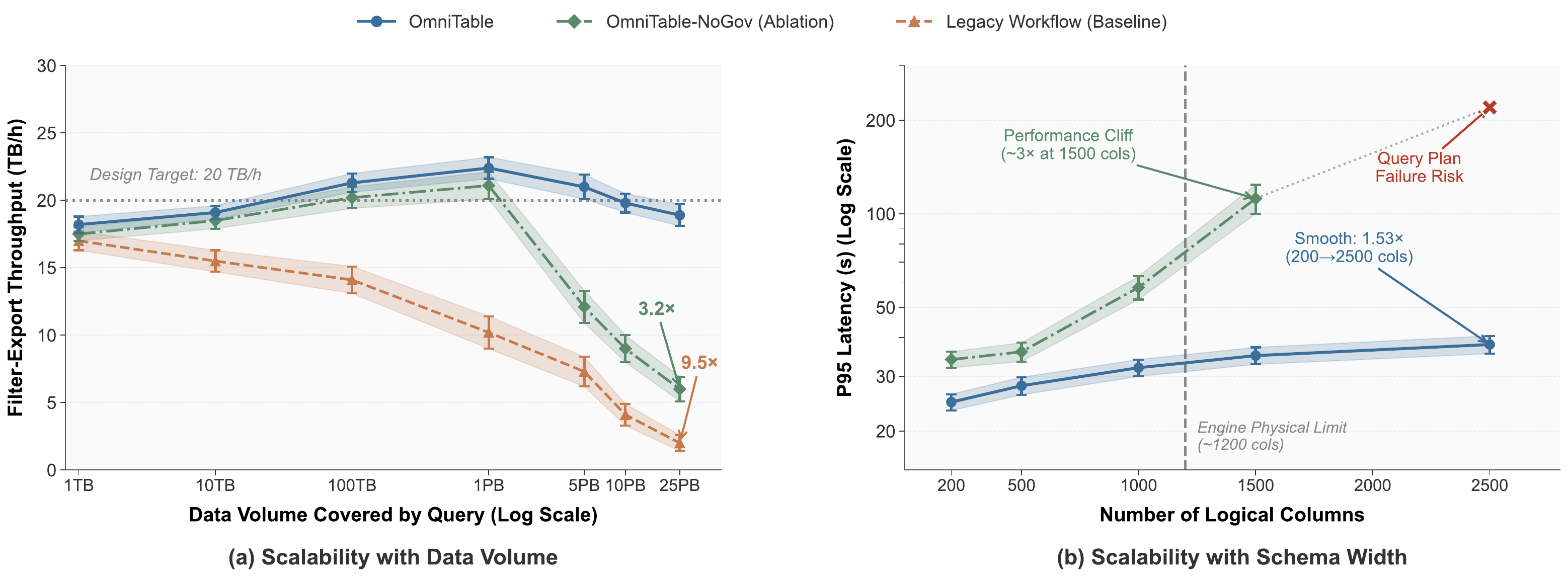}
  \caption{Scalability evaluation. (a)~Filter-export throughput vs.\ data volume (1\,TB--25\,PB). OmniTable maintains 18--23\,TB/h; OmniTable-NoGov degrades beyond 1\,PB; Legacy Workflow drops to ${\sim}$2\,TB/h at 25\,PB (9.5$\times$ slower). (b)~P95 query latency vs.\ number of logical columns (200--2500). OmniTable's column splitting transparently crosses the ${\sim}$1200-column engine limit (1.53$\times$ over full range); NoGov hits a ${\sim}$3$\times$ performance cliff at 1500 columns and query failures beyond.}
  \Description{Two scalability charts showing export throughput over data volume and P95 latency over logical column count.}
  \label{fig:scalability}
\end{figure*}

This section evaluates query performance as data volume and schema width grow, validating the effectiveness of autonomous governance (compaction, row splitting, column splitting). We compare OmniTable, OmniTable-NoGov, and Legacy Workflow.

\textbf{Experimental Design.} Two controlled experiments on \texttt{web}: (1)~\emph{Data volume scalability}: a fixed 15-column filtered export query with 8 predicates on 12 feature columns; data coverage varied from 1\,TB to 25\,PB by selecting different batch combinations. (2)~\emph{Schema width scalability}: fixed ${\sim}$2\,PB coverage; logical columns increased from 200 to 2500 by adding lightweight features; a fixed 10-column interactive filter query returning 1000 rows, ensuring the query column set is constant regardless of total column count.

\textbf{Data Volume Scalability (Figure~\ref{fig:scalability}a).} OmniTable maintained stable throughput of 18--23\,TB/h from 1\,TB to 25\,PB, due to the synergy of three mechanisms: small-file compaction keeps files in the 256\,MB--1\,GB optimal range for sequential I/O; batch pruning via Catalog metadata skips irrelevant physical tables; and row splitting bounds per-partition size for linear parallelism scaling.

OmniTable-NoGov shows comparable throughput below 1\,PB but declines sharply beyond: ${\sim}$10\,TB/h at 5\,PB (3.2$\times$ gap vs.\ OmniTable) and ${\sim}$5\,TB/h at 25\,PB, due to accumulated small files degrading I/O and unsplit partitions creating long-tail tasks. Legacy Workflow degrades most severely (${\sim}$2\,TB/h at 25\,PB, 9.5$\times$ slower than OmniTable), because cross-table queries require runtime multi-way JOINs with shuffle costs scaling linearly with data volume.

\textbf{Schema Width Scalability (Figure~\ref{fig:scalability}b).} With column splitting enabled, OmniTable's P95 latency increased from ${\sim}$25\,s to ${\sim}$38\,s (1.53$\times$) as columns grew from 200 to 2500---notably crossing MaxCompute's ${\sim}$1200-column physical limit with no performance discontinuity. Column splitting automatically migrates low-frequency columns to auxiliary tables based on access frequency; since the fixed query involves only high-frequency columns residing in the main table, it completes via single-table scan without JOINs.

OmniTable-NoGov shows similar latency below 800 columns but exhibits a sharp degradation near the engine limit: P95 rises to ${\sim}$55\,s at 1000 columns and ${\sim}$110\,s at 1500 columns (${\sim}$3$\times$ OmniTable). Beyond 1200 columns, some queries fail entirely due to exceeding the physical column limit, requiring manual table redesign.

\textbf{Summary.} OmniTable demonstrates stable performance across four orders of magnitude in data volume (1\,TB $\rightarrow$ 25\,PB) and exceeding the engine's physical column limit by ${\sim}$2$\times$ (200 $\rightarrow$ 2500 columns). The ablation study confirms that autonomous governance is indispensable: without it, throughput degrades by $>$3$\times$ at PB scale, and schema growth causes performance cliffs and query failures near engine limits.

\subsection{Engine Intelligence and Elasticity}
\label{sec:engine}

\begin{figure}[t]
  \centering
  \includegraphics[width=\columnwidth]{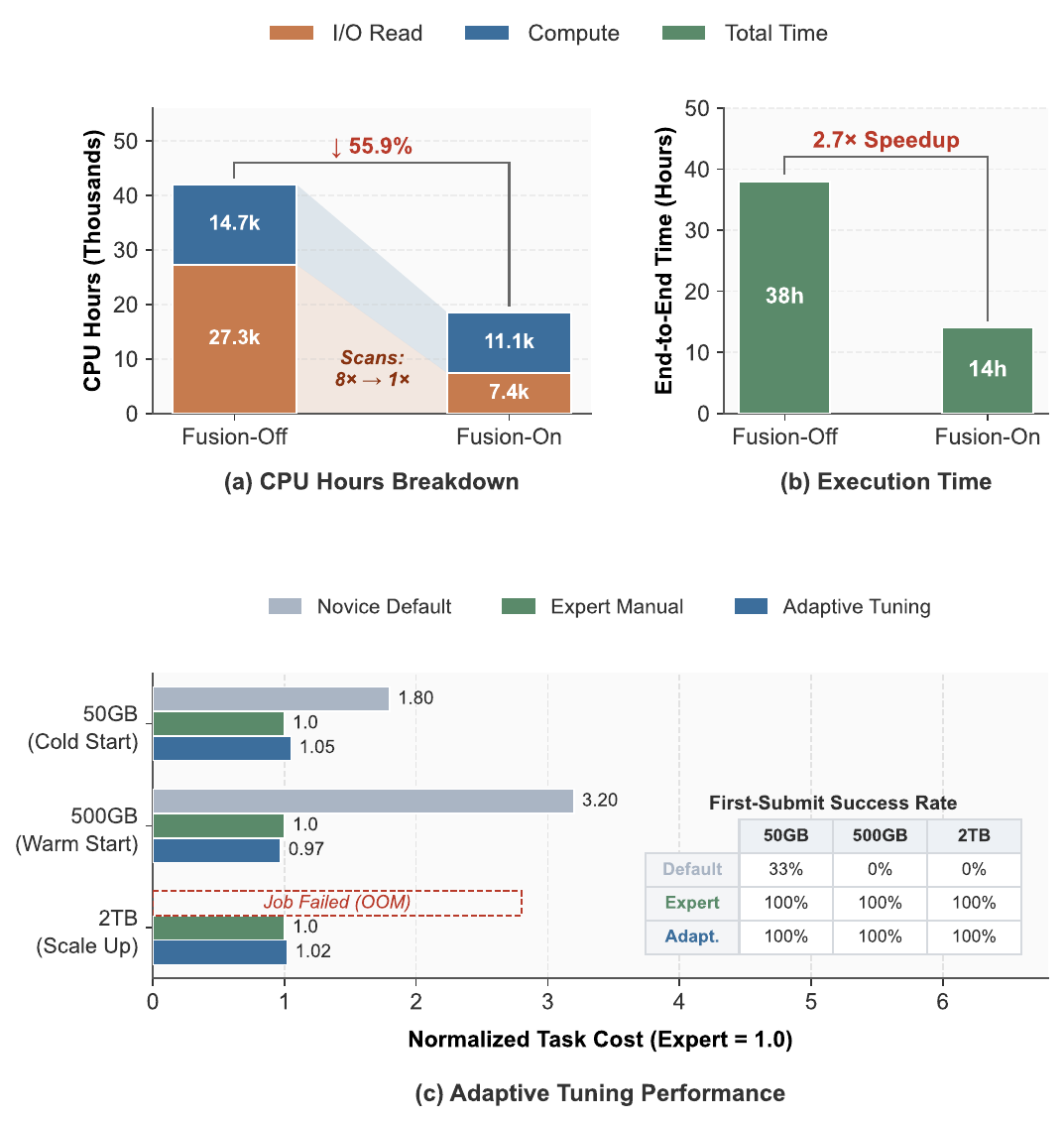}
  \caption{Execution evaluation. (a)~CPU Hours breakdown: Fusion-On reduces total CPU Hours by 55.9\% (scans: 8$\times$ $\rightarrow$ 1$\times$). (b)~End-to-end time: 2.7$\times$ speedup (38\,h $\rightarrow$ 14\,h). (c)~Adaptive tuning normalized task cost (Expert\,=\,1.0) across batch sizes; Default fails (OOM) at 2\,TB. Inset: first-submit success rates.}
  \Description{Three execution charts showing CPU hour breakdown, end-to-end time, and adaptive tuning cost across batch sizes.}
  \label{fig:engine}
\end{figure}

This section quantifies three key mechanisms of the execution engine through job-level controlled experiments on \texttt{web}. All results are medians of three replicates; resource consumption is measured in CPU Hours (cores $\times$ hours) including retry overhead.

\textbf{Experiment 1: Operator Fusion.} We selected 8 CPU/Spark features sharing \texttt{parsed\_text} as input on a Common Crawl batch (${\sim}$2.5\,PB, 300B+ records), covering text length, language detection, n-gram repetition, URL-content ratio, digit proportion, garbage-character flags, and content MD5. \emph{Fusion-Off} generates 8 independent tasks each scanning \texttt{parsed\_text}; \emph{Fusion-On} merges them into a single task producing 8 columns in one scan. As shown in Figure~\ref{fig:engine}(a--b), Fusion-On reduces total CPU Hours by 55.9\% (42K $\rightarrow$ 18.5K), with I/O read dropping from 27.3K to 7.4K CPU Hours (8$\times$ $\rightarrow$ 1$\times$ scans). End-to-end time decreases from 38\,h to 14\,h (2.7$\times$ speedup).

\textbf{Experiment 2: UDF-Level Failover.} We backfilled the fastText~\cite{Joulin17} math-recall feature \texttt{math\_recall\_v4} on a 500\,GB batch (${\sim}$600M records) containing ${\sim}$31K anomalous records (0.005\%) that trigger OOM due to extreme text length or encoding errors. We compared three configurations (Table~\ref{tab:failover}): Failover-On completes 99.995\% of records in one pass (${\sim}$6.2\,h, zero manual intervention), logging 31,247 anomalies to an error table. Failover-Off fails entirely. Legacy Workflow requires 3 manual investigate-remove-resubmit cycles (${\sim}$52\,h total, including ${\sim}$18\,h manual effort). This demonstrates that even at 0.005\% anomaly rate, task-level failure granularity causes highly asymmetric costs; record-level isolation eliminates this.
\begin{table}[t]
  \centering
  \caption{UDF-level failover comparison on \texttt{math\_recall\_v4} (500\,GB, ${\sim}$600M records, ${\sim}$31K anomalous at 0.005\%).}
  \label{tab:failover}
  \resizebox{\columnwidth}{!}{%
  \begin{tabular}{lccccc}
    \toprule
    \textbf{Config.} & \textbf{Success} & \textbf{Records OK} & \textbf{Anomalies} & \textbf{Time} & \textbf{Manual} \\
    \midrule
    Legacy      & 0\%$\!\to\!$100\% & 600M    & ${\sim}$31K (removed) & ${\sim}$52\,h & 3 rounds \\
    Failover-Off & 0\% (failed)      & 0       & ---                    & Failed        & $\geq$1  \\
    Failover-On  & 100\% (one-pass)  & 599.97M & 31,247 (logged)       & ${\sim}$6.2\,h & 0        \\
    \bottomrule
  \end{tabular}%
  }
\end{table}

\textbf{Experiment 3: Adaptive Tuning.} We backfilled \texttt{quality\_\allowbreak score} (BERT-based~\cite{Devlin19}, ${\sim}$350\,MB model, sensitive to \texttt{\seqsplit{memoryOverhead}}) on three batch sizes: 50\,GB, 500\,GB, and 2\,TB. We compared Default (Spark defaults), Adaptive (OmniTable auto-tuning), and Expert (manual tuning by experienced engineers). The metric is Normalized Task Cost $=$ CPU Hours $\times$ (1 + Retry Count), with Expert normalized to 1.0.

As shown in Figure~\ref{fig:engine}(c), Adaptive achieves 100\% first-submit success across all sizes with cost within 5\% of Expert. Default fails at $\geq$500\,GB (OOM). At 500\,GB, Adaptive (0.97) slightly outperforms Expert (1.0) because HBO interpolation produces tighter parameters than human experts' conservative margins. The progressive strategy uses heuristic rules for cold starts (increasing \texttt{memoryOverhead} based on data profiling) and HBO for warm starts.

\textbf{Combined Effect.} In the SFT scenario of \S5.2 (feature backfill 7.8\,d saving), the three mechanisms contribute complementary savings: operator fusion saves ${\sim}$1.5\,d of I/O overhead; UDF fault tolerance avoids ${\sim}$2--3\,d of investigate-retry cycles; adaptive tuning eliminates ${\sim}$1--2\,d of parameter-related failures. These partially overlap, totaling ${\sim}$5--6.5\,d; the remainder comes from scheduling improvements.

\subsection{Hybrid-Accelerated Data Exploration}
\label{sec:exploration_eval}

\begin{figure*}[t]
  \centering
  \includegraphics[width=\textwidth]{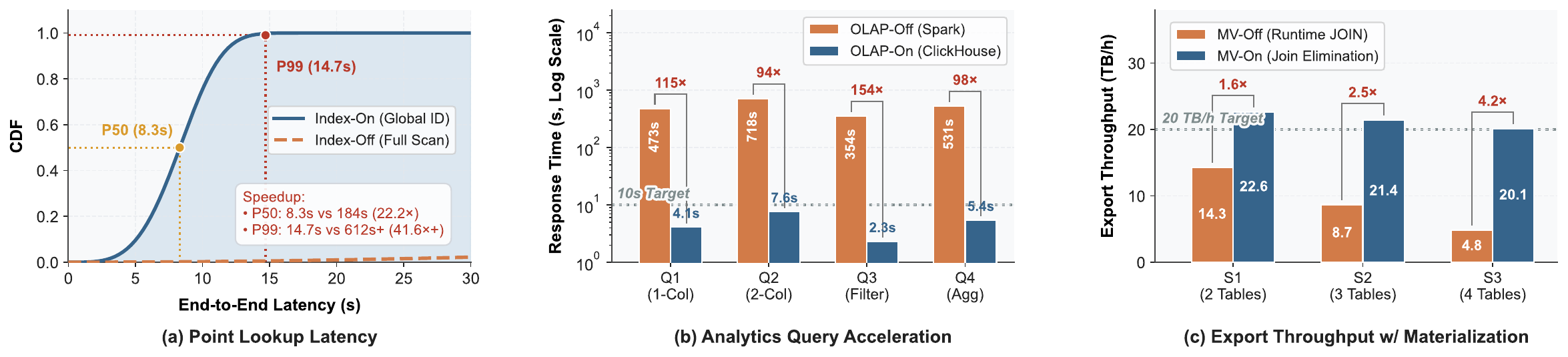}
  \caption{Hybrid acceleration for data exploration. (a)~Point lookup latency CDF: global ID index achieves P50\,=\,8.3\,s vs.\ 184\,s full scan (22.2$\times$), P99\,=\,14.7\,s vs.\ 612\,s+ (41.6$\times$+). (b)~Analytical query response time (log scale): OLAP offloading to ClickHouse achieves 94--154$\times$ speedup, all under 10\,s. (c)~Filter-export throughput: materialized views eliminate runtime JOINs, achieving 1.6--4.2$\times$ speedup and sustaining ${\geq}$20\,TB/h across scenarios.}
  \Description{Three charts showing point lookup latency, analytical query response time, and filter-export throughput with hybrid acceleration.}
  \label{fig:exploration}
\end{figure*}

This section evaluates the three complementary acceleration technologies in the Exploration Service on \texttt{web} (25\,PB, 300B+ records, 800+ logical columns, 6 physical tables). All queries were replayed at concurrency 20; results are medians of three runs.

\textbf{Point Query Performance (Figure~\ref{fig:exploration}a).} With the global ID index enabled, single-record lookups by \texttt{\_ai\_unique\_id\_} achieve P50 latency of 8.3\,s and P99 of 14.7\,s. Without the index, the system must scan all physical tables that may contain the target ID, yielding P50 of 184\,s and P99 of 612\,s+---a speedup of 22.2$\times$ (P50) to 41.6$\times$+ (P99). Even at P99, a complete logical row (800+ columns) is returned within 15\,s, transforming anomaly backtracking from batch submission to interactive querying.

\textbf{Analytical Query Acceleration (Figure~\ref{fig:exploration}b).} OLAP offloading to ClickHouse~\cite{Alexyev24} achieves 94--154$\times$ speedup across four representative queries (single-column COUNT, two-column GROUP BY, filtered COUNT, multi-column aggregation), reducing response times from minutes to under 10\,s. ClickHouse's sparse index pruning yields the highest speedup on filtered queries. Coverage spans ${\sim}$30 high-frequency columns; unsynchronized queries fall back to Spark.

\textbf{Filter-Export Acceleration (Figure~\ref{fig:exploration}c).} Background materialized views show increasing benefit as queries span more physical tables. S1 (5-column filter, 2 tables): MV-Off achieves 14.3\,TB/h (single JOIN, acceptable shuffle cost); MV-On reaches 22.6\,TB/h (1.6$\times$). S2 (10 columns, 3 tables): MV-Off drops to 8.7\,TB/h due to cascading JOINs; MV-On maintains 21.4\,TB/h (2.5$\times$). S3 (15 columns, 4 tables): MV-Off falls to 4.8\,TB/h, well below the 20\,TB/h target; MV-On rewrites the query as a single-table scan, sustaining 20.1\,TB/h (4.2$\times$). Materialized view benefit is positively correlated with the number of eliminated JOINs; typical ablation experiment queries (10--15 feature columns) fall in the S2--S3 range, making materialized views essential for meeting the throughput target.

\textbf{Synergistic Effect.} The three technologies cover complementary query patterns: ID index for point lookups (second-level latency), OLAP offloading for aggregation analysis (second-level response), and materialized views for multi-column filtered exports (${\geq}$20\,TB/h). The query optimizer selects the optimal path automatically. Together, they reduce the data filtering phase in the SFT scenario (\S5.2) from 2.5 days to ${\sim}$0.3 days, covering the full query chain from anomaly identification through distribution analysis to data export.

\section{Lessons Learned}

\noindent\textbf{Record-level fault tolerance is non-negotiable.} Before UDF-level isolation, 30--40\% of engineering hours were spent on failure investigation. A single malformed record among billions routinely failed entire multi-TB jobs. The 3--5\% execution overhead of per-record wrapping eliminated our most expensive category of human intervention.

\noindent\textbf{Autonomous governance is existential, not optional.} We initially deferred background compaction and splitting as ``nice-to-have.'' Within three months, small-file accumulation degraded query latency by 3--5$\times$ and column growth hit engine limits. Without continuous physical layout evolution, a PB-scale continuously-written system becomes unusable within weeks.

\noindent\textbf{Adoption beats capability if engineers won't change habits.}
Algorithm engineers lived and breathed SQL: their instinct was
to spin up yet another table and write yet another query. Asking them to
abandon that habit upfront was a losing battle. Instead, we met them where
they were: the logical wide table speaks plain SQL; a feature template library
let them register a new feature in minutes rather than wiring a new pipeline;
and lineage records are queryable tables, so auditing felt no different from
\texttt{SELECT}ing a result. Only after these pain points disappeared did the
deeper benefits---automated dependency resolution, heterogeneous routing,
autonomous governance---become visible. A system that demands a workflow
change before delivering value will be worked around, not adopted.

\section{Related Work}

\textbf{Lakehouse Architecture.} Delta Lake~\cite{Armbrust20}, Apache Iceberg~\cite{Blue17}, and Apache Paimon~\cite{ASF23a} provide ACID, time travel, and schema evolution on data lakes; OmniTable can be viewed as a reusable layer above such storage. Lakehouse systems are largely \emph{table-centric} and do not treat a \emph{column/UDF dependency DAG} and \emph{operator-aware CPU/GPU routing} as first-class lifecycle objects. OmniTable targets these LLM-specific needs via Catalog-managed feature metadata, dependency closure, routing, and autonomous layout optimization~\cite{Armbrust21}.

\textbf{Feature Store.} Feast~\cite{Feast19}, Tecton~\cite{Tecton20}, and Hopsworks~\cite{Dowling23} manage ML features with emphasis on low-latency online serving of pre-computed structured features. OmniTable addresses a different problem: PB-scale backfill of UDF-defined features over raw unstructured text with CPU/GPU routing, governed across the full data lifecycle. These settings are sufficiently distinct that a shared benchmark does not exist; quantitative comparison would conflate compute efficiency with orchestration overhead.

\textbf{Data Orchestration and Transformation.} dbt~\cite{dbt16} and Apache Airflow~\cite{ASF15} operate at table and task granularity without domain-specific automation (dependency resolution, heterogeneous routing, UDF-level fault tolerance, adaptive tuning). OmniTable elevates operations to rows and columns on a logical wide table with these capabilities built in.

\textbf{LLM Data Processing Frameworks.} Data-Juicer~\cite{Chen24}, RedPajama~\cite{Together23}, Dolma~\cite{Soldaini24}, and SmallPond~\cite{DeepSeek25} provide operator libraries for LLM data cleaning but follow the pipeline paradigm without unified cross-dataset management. OmniTable operates at a different layer: it provides orchestration, governance, and storage abstraction \emph{above} operator libraries, complementing rather than competing with them---their operators can be directly registered as OmniTable UDFs.

\textbf{ML Lifecycle and Data Management.} TFX~\cite{Baylor17}, Polyzotis et al.~\cite{Polyzotis18}, Whang et al.~\cite{Whang23}, and Snorkel~\cite{Ratner17} motivate the data-centric perspective but do not provide the unified storage abstraction, declarative feature backfill, or autonomous governance that OmniTable contributes for PB-scale LLM scenarios.

\section{Conclusion and Future Work}

OmniTable addresses heterogeneity, scale, agility, and traceability in PB-scale LLM data governance via \emph{Logical Unification, Physical Separation}. It combines a unified wide-table abstraction, declarative feature lifecycle management, an adaptive execution engine, and hybrid acceleration to reduce exploration latency from minutes to seconds while sustaining ${\geq}$20\,TB/h export throughput. In production, OmniTable manages over 35\,PB of LLM training data, reducing the governance cycle from ${\sim}$14 days to ${\sim}$2.5 days (5.6$\times$).

Future directions include intelligent data recommendation to close the Data-Centric AI loop~\cite{Ratner17, Whang23}, batch-stream unification for near-real-time feature computation~\cite{Carbone15}, and synthetic data governance with generation lineage tracking.

\clearpage

\balance
\bibliographystyle{ACM-Reference-Format}
\bibliography{sample}

\end{document}